\documentclass[a4paper,fleqn,usenatbib]{mnras}

\usepackage{mathptmx}

\usepackage[T1]{fontenc}
\usepackage{ae,aecompl}

\usepackage{graphicx}	
\usepackage{amsmath}	
\usepackage{amssymb}	

\newcommand{\pcm}{cm$^{-2}$}
\newcommand{\xmm}{\textit{XMM-Newton}}
\newcommand{\nustar}{\textit{NuSTAR}}
\newcommand{\swift}{\textit{Swift}}
\newcommand{\delcstat}{$\Delta$C-stat}

\title[A Variable Outflow in the NGC 300 PULX]{Evidence for a Variable Ultrafast Outflow in the Newly Discovered Ultraluminous Pulsar NGC 300 ULX-1}

\author[P. Kosec et al.]{P. Kosec$^{1}$\thanks{E-mail: pk394@cam.ac.uk},
C. Pinto$^{1}$,
D. J. Walton$^{1}$,
A. C. Fabian$^{1}$,
M. Bachetti$^{2}$,\newauthor 
M. Brightman$^{3}$,
F. F\"urst$^{4}$
and B. W.~Grefenstette$^{5}$
\\
$^{1}$Institute of Astronomy, Madingley Road, CB3 0HA Cambridge, UK\\
$^{2}$INAF-Osservatorio Astronomico di Cagliari, via della Scienza 5, I-09047 Selargius (CA), Italy\\
$^{2}$Cahill Center for Astrophysics, California Institute of Technology, 1216 East California Boulevard, Pasadena, CA 91125, USA\\
$^{4}$European Space Astronomy Centre (ESAC), Science Operations Department, 28692 Villanueva de la Ca\~nada, Madrid, Spain\\
$^{5}$Space Radiation Lab, 1200 E. California Blvd, California Institute of Technology, Pasadena, CA 91125, USA
}

\date{Accepted 2018 June 12. Received 2018 June 12; in original form 2018 March 5}

\pubyear{2018}

\begin{document}
\label{firstpage}
\pagerange{\pageref{firstpage}--\pageref{lastpage}}
\maketitle

\begin{abstract}

Ultraluminous pulsars are a definite proof that persistent super-Eddington accretion occurs in nature. They support the scenario according to which most Ultraluminous X-ray Sources (ULXs) are super-Eddington accretors of stellar mass rather than sub-Eddington intermediate mass black holes. An important prediction of theories of supercritical accretion is the existence of powerful outflows of moderately ionized gas at mildly relativistic speeds. In practice, the spectral resolution of X-ray gratings such as RGS onboard XMM-Newton is required to resolve their observational signatures in ULXs. Using RGS, outflows have been discovered in the spectra of 3 ULXs (none of which are currently known to be pulsars). Most recently, the fourth ultraluminous pulsar was discovered in NGC 300. Here we report detection of an ultrafast outflow (UFO) in the X-ray spectrum of the object, with a significance of more than 3$\sigma$, during one of the two simultaneous observations of the source by \xmm\ and \nustar\ in December 2016. The outflow has a projected velocity of 65000 km/s (0.22c) and a high ionisation factor with a log value of 3.9. This is the first direct evidence for a UFO in a neutron star ULX and also the first time that this its evidence in a ULX spectrum is seen in both soft and hard X-ray data simultaneously. We find no evidence of the UFO during the other observation of the object, which could be explained by either clumpy nature of the absorber or a slight change in our viewing angle of the accretion flow.

\end{abstract}




\section{Introduction}

Ultraluminous X-ray sources (hereafter ULXs) are point-like, non-nuclear objects with an isotropic X-ray luminosity exceeding the Eddington luminosity of standard stellar (10 $M_{\odot}$) black holes. Initially, it was thought that they could be powered by intermediate mass (100s-10000s $M_{\odot}$) black holes accreting at sub-Eddington rates. This could still be true for a minority of these objects, especially the more luminous ones such as ESO 243-49 HLX-1 \citep{Farrell+09, Webb+12}. 

However, recent evidence suggests that a majority of the population might in fact be formed of supercritical stellar mass accretors. Most ULXs have been shown to have a very low high-energy cutoff, seen above 5 keV in \xmm\ and especially \nustar\ spectra \citep{Bachetti+13, Walton+14}. This is inconsistent with standard models of sub-Eddington accretion, where a significant fraction of luminosity is expected above 10 keV, as observed in X-ray binaries and active galactic nuclei (AGN). More recently, it was discovered that 4 ULXs are powered by neutron stars thanks to their pulsating behaviour \citep{Bachetti+14,Furst+16,Israel+17a, Israel+17b,Carpano+18b}. The discovery of a likely cyclotron feature in the spectrum of M51 ULX8 \citep{Brightman+18} suggests that this source also harbours a neutron star. The existence of pulsating ultraluminous X-ray sources (PULXs) is a definite proof that supercritical accretion occurs in at least a fraction of the ULX population. At the same time, there is currently no strong evidence against the possibility that most other ULXs might in fact be neutron stars \citep{Pintore+17,Walton+18}.

Models of super-Eddington accretion \citep{Shakura+73, Poutanen+07}, as well as simulations \citep{Narayan+17} predict a geometrically and optically thick disk around the compact object. An optically thin, evacuated funnel is formed along the rotation axis of the object. The opening angle of the funnel is thought to be dependent on the Eddington ratio of the object \citep{King+09}. The excess radiation pressure is converted into kinetic energy of a fraction of the accreted matter, which is launched away from the system in the form of a mildly relativistic (0.1-0.3c) outflow. Simulations show that these outflows tend to cover a wide solid angle ($\sim$0.5) and are clumpy \citep{Takeuchi+13}. For a current idea of the structure of the system, see Fig. 13 of \citet{Pinto+17}. In addition, recent simulations also predict the existence of collimated jets at relativistic velocities. The accretion geometry is even more complicated in the case of neutron star ULXs, due to the presence of strong magnetic fields which heavily affect the accretion flow near the central object. Consequently, an accretion column is formed, and is likely the source of most of the hard X-ray radiation \citep{Mushtukov+17,Walton+18}.

Such a configuration necessarily produces very different observational signatures when viewed under different inclination angles. At low inclination angles (looking down the evacuated funnel), the observer sees the innermost and hottest regions of the accretion flow, which produce the hardest radiation (mostly hard X-rays). At increasing inclination angles, the outflow begins to obscure the innermost regions, softening the overall spectrum of the object. At high inclination, the outflow becomes optically thick and the spectrum of the source is therefore very soft, possibly producing the spectrum of an ultraluminous supersoft source \citep{Urquhart+16}.

Until recently, the ultrafast outflows in ULXs were not proven observationally because they are unresolvable with the current generation of X-ray CCD instruments. However, soft residuals were spotted in CCD spectra of many ULXs \citep[][and references therein]{Stobbart+06}. \citet{Middleton+15b} showed that in the case of NGC 1313 X-1, these residuals vary in time and anticorrelate with the luminosity of the object. This result suggested that they are a signature of outflows and not of relativistic reflection or diffuse emission from nearby star formation. \citet{Pinto+16, Pinto+17} used high resolution X-ray gratings onboard \xmm\ to detect winds in 3 ULXs: NGC 1313 X-1, NGC 5408 X-1 and NGC 55 ULX. \citet{Kosec+18} studied 10 other ULXs and found that they show signatures similar to those of the 3 ULXs with outflows, but the current data quality was not sufficient to claim any significant detections. Instead, the first X-ray detection of a jet in a ULX (with a projected velocity of 0.34c) was achieved in NGC 5204 X-1. Other ULXs have also been shown to harbour jets thanks to radio studies \citep{Middleton+13,Cseh+15}.

At the moment, none of the neutron star ULXs are known to possess ultrafast outflows. This is probably caused by observational difficulties - out of the 3 PULXs known by the end of 2017, only 1 is suitable for a high-resolution X-ray analysis with the current generation of instruments. The remaining 2 are either too distant or located in a crowded field.

However, in January 2018, the 4th ultraluminous pulsar was discovered in NGC 300 \citep{Carpano+18a, Carpano+18b} and named NGC 300 ULX-1. The object was originally mis-identified as a supernova and named SN2010da \citep{Monard+10,Elias-Rosa+10} but later identified as a likely supergiant B[e] high-mass X-ray binary \citep{Binder+11,Lau+16,Villar+16}. It has only recently been observed at ULX luminosities, and is experiencing an extremely fast spin-up \citep{Kennea+18}, having spun-up from a period of 45 s to just 20 s in less than 2 years. Furthermore, \citet{Walton+18b} discovered a potential cyclotron resonant scattering feature in the hard X-ray spectrum of the object.

In this work, we perform a rigorous, in-depth analysis of the coordinated \xmm\ and \nustar\ data from December 2016, searching for spectral signatures of an ultrafast outflow in this system. We use simple Gaussian line scans as well as automated searches with physical models of outflowing photoionised absorbers and quantify the statistical significance of detections with Monte Carlo simulations. We detect a variable UFO with a projected velocity of 0.22c in the spectrum of the object with a significance of more than 3$\sigma$.

The structure of this paper is as follows. Section \ref{sec:Data} contains information about the observations of the source and data reduction. We present the methods and results in section \ref{sec:Results} and discuss their implications in section \ref{sec:Discussion}. Finally, section \ref{sec:Conclusions} summarizes our findings.

\section{Observations and Data Reduction}
\label{sec:Data}

We used data from \xmm\ and \nustar. There are multiple observations of the object by the \swift\ satellite, but its spectral resolution is not sufficient for a search for ultrafast wind in a ULX.

As of February 2018, there are two simultaneous observations of the source by \nustar\ and \xmm\ taken very shortly within each other in December 2016. The \xmm\ exposures are separated by a 10 hour window because of the satellite's orbit. The first one is about 140 ks of raw time long, the exposure of the second one is about 80 ks of raw time. \nustar\ observed the object at the same time, for a total duration of about 320 ks ($\sim$160 ks exposure time). We extracted spectra for each of the \xmm\ observations separately, without any stacking during the analysis, and also extracted 2 \nustar\ spectra from time intervals that are simultaneous with \xmm\ observations to take into account possible variability of the object (albeit its count rate, $\sim$0.9 counts/s with the PN instrument, is very similar during both observations). Further info about the exposures is shown in Table \ref{obsdata}.

We obtained the object distance of 2 Mpc by averaging newer measurements on the NED database.

\begin{table}
	\centering
	\caption{Log of the observations used in this work. Column (1) lists the instrument. Column (2) contains the observation ID and (3) the start time of the exposure. Column (4) lists the clean exposure time in ks.}
	\label{obsdata}
	\begin{tabular}{cccc} 
		\hline
		Instrument&Obs ID&Start time&Exposure$^{a}$\\
		 & & & ks \\
		(1)  & (2) & (3) & (4) \\
		\hline
		\xmm &0791010101&2016-12-17 08:52&134/96/120\\
		\xmm &0791010301&2016-12-19 08:44&77/46/65\\
		\nustar &30202035002&2016-12-16 15:31&163\\
		\hline
	\end{tabular}
	$^{a}$\xmm\ exposures are listed as RGS/PN/MOS exposure.
\end{table}

\subsection{XMM-Newton}

All the \xmm\ \citep{Jansen+01} data was downloaded from the XSA archive and reduced using a standard pipeline with SAS v16, CalDB as of January 2018. We use data from all X-ray instruments onboard \xmm: the European Photon Imaging Camera (EPIC): PN \citep{Struder+01} and MOS \citep{Turner+01} CCDs, and the Reflection Grating Spectrometer (RGS) detectors \citep{denHerder+01}. We filter all high-background periods, with a threshold of 0.5 counts/sec for PN data, 0.25 counts/sec for MOS and RGS data. Only events of PATTERN <= 4 (single/double) were accepted for PN data, and events of PATTERN<=12 were included in MOS datasets, according to standard procedures. The source regions for PN and MOS 1,2 were circles centred on the ULX with a radius of 40 arcsec. Unfortunately, the position of the ULX is right on a chip gap of the PN chip, so the data might be partly compromised. This is why we also use MOS 1 and 2 data which are unaffected by this issue. However, the agreement between PN and MOS spectra is very good, hence the PN chip gap position is probably not a problem. The background regions were circles located in the same region of the chip (and avoiding the copper ring and out of time events on the PN chip), with a radius of 50 arcsec.

\xmm\ was not pointed directly at the object and therefore we cannot use the default selection of source and background regions for the RGS detectors. The default source region is pointed on the X-ray binary NGC 300 X-1 (at a distance of about 1.1 arcmin from NGC 300 ULX-1). We chose a source region centred on the position of NGC 300 ULX-1, with a size such that it covers 90 per cent of the instrument PSF in the cross-dispersion direction. The background region is a rectangle shifted by 1.4 arcmin in the cross-dispersion direction, with a size that includes 95 percent of the PSF, avoiding the flux from NGC 300 X-1 and at the same time maximizing its area to constrain the background as well as possible. 
 
All reduced data were converted from the OGIP to the SPEX format using the Trafo\footnote{http://var.sron.nl/SPEX-doc/manual/manualse100.html} tool so they can be used by the SPEX fitting package. We grouped the PN, MOS 1 and MOS 2 data to at least 25 counts per bin and also rebinned the original EPIC channels by at least a factor of 3 using the SPECGROUP procedure. RGS data was binned by a factor of 3 directly in SPEX to oversample the spectral resolution by about a factor of 3. The spectral range used was: 0.3 to 10 keV for PN data, 0.3 to 9.8 keV for MOS, initially 7 \AA\ (1.8 keV) to 26 \AA\ (0.5 keV), and later only 7 to 20.5 \AA\ (0.6 keV) for RGS data, mostly limited by the background level and the calibration uncertainties of the instruments.

\subsection{NuSTAR}

We reduced the \nustar\ \citep{Harrison+13} data following standard procedures using the \nustar\ Data Analysis Software (NUSTARDAS, v1.8.0) and instrumental calibration files from CalDB v20171204. We first cleaned the data with NUPIPELINE, using the standard depth correction, which significantly reduces the internal high-energy background, and also removed passages through the South Atlantic Anomaly (using the settings SAACALC=3, TENTACLE=NO and SAAMODE=OPTIMIZED). Source and background spectra and instrumental responses were then produced for each of the two focal plane modules (FPMA/B) using NUPRODUCTS. Source products were extracted from circular regions of radius 40$''$, and background was estimated from larger regions of blank sky on the same detector as ULX-1. In order to maximise the signal-to-noise (S/N), in addition to the standard `science' (mode 1) data, we also extracted the `spacecraft science' (mode 6) data following the method outlined in \citet{Walton+16cyg}. In this case, mode 6 provides $\sim$10\% of the good \nustar\ exposure.

\section{Methods and Results}
\label{sec:Results}

In this section, we describe the methods we used to study the data and show the results of our analysis. Note that all the subsections except for the last one only use the data taken during the second \xmm\ observation of the object (ID 0791010301) alongside with simultaneous \nustar\ data. We also performed the same analysis on the second dataset (ID 0791010101) but find no evidence for ultrafast wind.

\subsection{Broadband modelling}
\label{broadbandfit}

We use the SPEX fitting package \citep{Kaastra+96} for spectral fitting, and C-statistics \citep{Cash+76} as there are not enough counts per bin in RGS data for a $\chi^{2}$ analysis. All the model parameters are checked extensively with a proper error search in case there are multiple minima in the C-stat function. The statistical error intervals are calculated at 1$\sigma$ level (68 per cent confidence).

We fit the broadband X-ray spectrum of the object with a phenomenological model which consists of a powerlaw (\textsc{po} in SPEX) plus a simple blackbody (\textsc{bb}) plus a color corrected blackbody (\textsc{mbb}). Similar spectral models have been used by \citet{Gladstone+09} and \citet{Walton+14} and describe the ULX continuum reasonably well. We also test alternative descriptions of continuum such as a disk blackbody component and find that the results of our analysis do not change significantly. All the emission is absorbed by Galactic and host galaxy (mostly) neutral interstellar medium (ISM) accounted for by the \textsc{hot} component \citep[see][for an example of similar usage of the model]{Kaastra+06}. The Leiden/Argentine/Bonn Survey of Galactic HI \citep{Kalberla+05} shows that there should be a column density of about $4 \times 10^{20}$~\pcm\ towards the host galaxy NGC 300.

We use all available data for this observation from RGS, PN, MOS (1 and 2) and FPM (A and B) detectors to constrain the broadband continuum as well as possible. A cross-normalisation constant is added for each instrument to account for small (<10 per cent) calibration uncertainties of each detector. The best-fit model is a powerlaw with a hard slope of $1.51^{+0.12}_{-0.21}$, a soft blackbody with a temperature of $0.231^{+0.008}_{-0.007}$ keV and a color-corrected blackbody with a temperature of $2.67\pm0.07$ keV. We obtain the ISM column density of $2.7^{+1.1}_{-1.3} \times 10^{20}$~\pcm. The fit statistics is C-stat=1545.87 for 1347 degrees of freedom. The unabsorbed 0.3 to 10 keV luminosity of the ULX during this epoch inferred by fitting the spectral components is about $3 \times 10^{39}$~erg~s$^{-1}$.

We note that the object appears to have an iron line at 6.7 keV in the MOS data, but this is not seen in either FPM nor PN data.

\subsection{Line search}

\begin{figure*}
	\includegraphics[width=\textwidth]{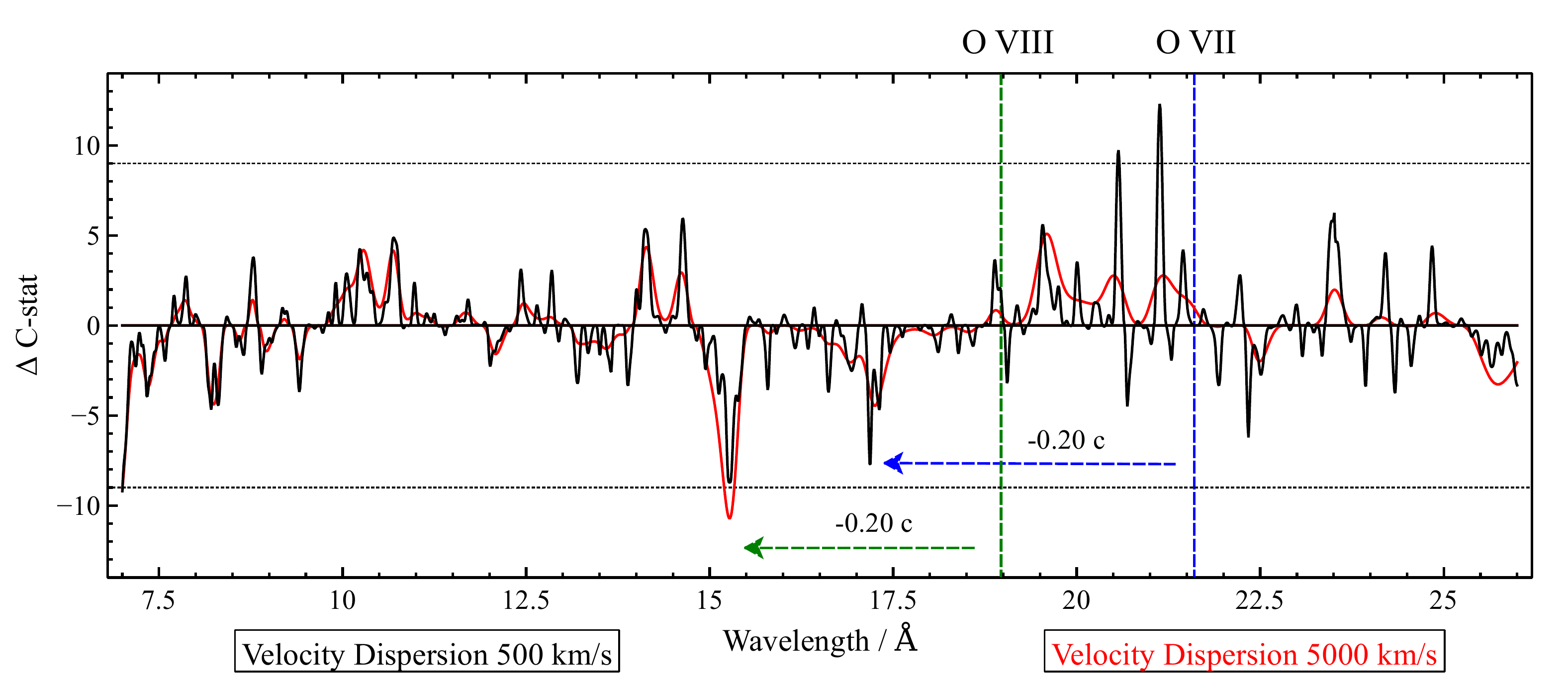}
    \caption{Gaussian line search results of RGS data from the second \xmm\ observation (0791010301). The black curve shows the line search using a velocity dispersion of 500 km/s, while the red curve shows the search assuming 5000 km/s. The Y axis is defined as $\Delta$C-stat times the sign of normalisation of the line to show the difference between absorption and emission features. Horizontal dotted lines show the values of $\Delta$C-stat=9 and -9. Position of the rest-frame wavelength of emission lines of the O VIII and O VII ions (taking the wavelength of the resonant line) is shown with green and blue dashed lines. Green and blue dashed arrows show the potential blueshift of the absorber if the potential features are produced by O VII and O VIII absorption.}
    \label{RGS_Gaussian}
\end{figure*}

Initially, we performed a simple Gaussian scan using the RGS data only. We aimed to find any possible narrow absorption and emission features that an outflow might imprint on the continuum spectrum of NGC 300 ULX-1. We used the EPIC broadband coverage to anchor the spectral continuum outside the RGS band, between 0.3 and 10 keV with the exception of the 7 to 26 \AA\ region (approximately 0.5 to 1.8 keV).

The scan is done by adding a Gaussian line with a fixed energy/wavelength to the continuum model. We also fix its width/turbulent velocity to avoid too large line widths and also to ensure a simple C-stat parameter space. The model is then fitted for the line normalisation, which can be either positive or negative (to allow for both emission and absorption features). The statistical fit improvement \delcstat\ is saved. Afterwards, the line is moved to another fixed energy in a grid. The grid ideally slightly oversamples the energy resolution of the instrument. In the case of RGS, we choose steps of 300 km/s, which works out to about 2000 steps across the 7 to 26 \AA\ wavelength range. To avoid unconstrained fits, we keep the continuum parameters frozen with the exception of the overall normalization of the broadband model (achieved by coupling individual component normalizations). This can be repeated for different line widths/turbulent velocities. 

The statistical significance of detection of a feature can be approximately inferred from the \delcstat\ fit improvement - this is a fit for 1 additional free parameter (the line normalisation) compared with the original continuum fit, hence an improvement of \delcstat$=9$ would give a significance of 3$\sigma$. This approach however ignores the large amount of trials performed to find features across the whole energy grid (the look-elsewhere effect) and so it overestimates the real significance. The real significance of a feature must be obtained by running Monte Carlo simulations which can be very computationally expensive. Therefore, in this section we will use the \delcstat\ value as a proxy to the strength of the features, but this number should be taken very cautiously.

We perform the Gaussian line scan on the RGS data of the second XMM observation (ID 0791010301) of the object using 2 different turbulent velocities: 500 km/s and 5000 km/s. The results are shown in Fig. \ref{RGS_Gaussian}. We notice that there are no very strong features (\delcstat~$>15$) in the spectrum. However, there is an absorption signature at around 15.3 \AA\ (0.81 keV) that could originate in blueshifted oxygen VIII absorption. The rest-frame wavelength of this transition is 19 \AA\, so the corresponding blueshift would be around 0.2c. Curiously, there is another, weaker absorption feature at 17.2 \AA\ (0.72 keV), that, if produced by O VII ions, would give the absorber blueshift of around 0.2c as well (see Fig. \ref{RGS_Gaussian}), assuming that the resonant line at the rest-frame wavelength of 21.6 \AA\ would dominate the absorption. This motivates us to follow-up with a more rigorous search, using a physical model of the absorber. 

Additionally, there seem to be a few emission lines between 20 and 22 \AA\ in the line scan, however they are quite narrow and located in the energy band ($>$20 \AA) which is partially affected by the background. They are also not located on a rest-frame wavelength of any expected transition.

\subsection{Physical model search}
\label{Phys_model_search}

\begin{figure*}
	\includegraphics[width=\textwidth]{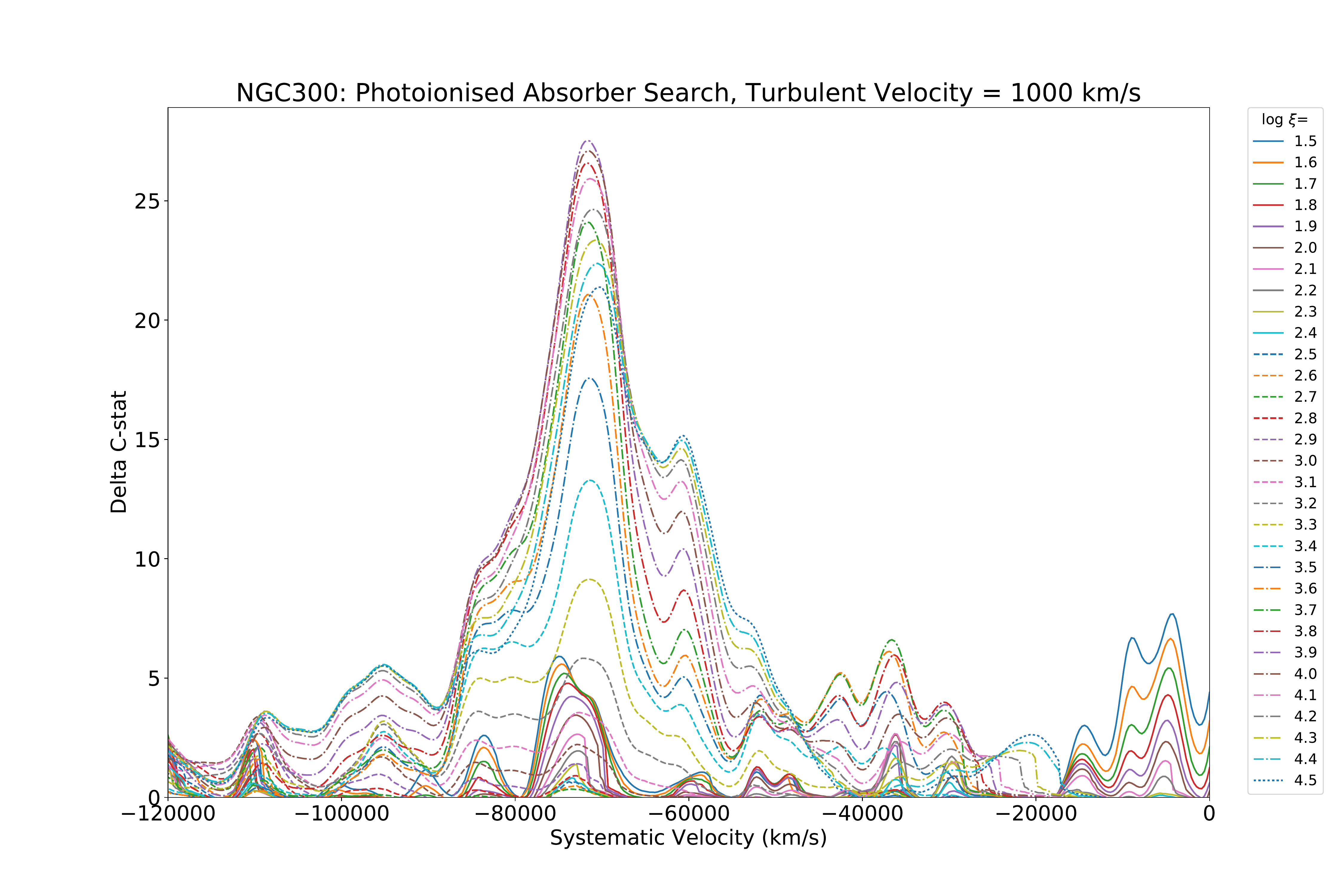}
    \caption{Automated scan of all available data using a photoionised absorber model \textsc{xabs} with a fixed turbulent velocity of 1000 km/s. The X-axis shows the systematic velocity of the absorber, while the Y-axis gives the \delcstat\ statistical fit improvement achieved by adding this component to the broadband continuum spectrum. Curves of different colors show automated searches with different values of the ionisation parameter which is kept frozen during the search.}
    \label{Finaldata_XABS}
\end{figure*}

Finding two features that could correspond to the same blueshift encourages us to try to fit the features with a physical photoionised absorption model. We use the \textsc{xabs} model in SPEX to describe the absorption of outflowing photoionised gas \citep{Steenbrugge+05} and find the best fit at a speed of 66000 $\pm$ 1000 km/s (0.220 $\pm$ 0.003 $c$), with a column density of about 10$^{21}$~\pcm, the ionisation parameter with a log value of 2.1 and a turbulent velocity of $\sim$2000 km/s. The \delcstat\ fit improvement is relatively modest at about 13.5.

However, upon closer inspection we also notice wiggles and features in the >2 keV PN spectrum. These could be unresolved absorption lines of higher energy transitions, not unlike those seen in some AGN, for example IRAS 13224-3809 \citep{Parker+17}. This suggests the possibility that the wind might actually be much more ionised than the best-fit RGS solution and that more information could be hidden in the higher energy CCD data. For this reason, we use all the data available: PN, MOS 1 and 2 data from the second \xmm\ observation of the object, plus we extract spectra from \nustar\ FPMA and FPMB instruments for the time duration of this observation.

To locate any possible solutions, we perform an automated scan with the \textsc{xabs} model, which is parametrised by the following main parameters: the column density and the systematic velocity of the absorber, the ionization state of the gas, and the turbulence velocity of the gas. We use the broadband continuum obtained in section \ref{broadbandfit}, which in this case must be completely freed (as we anticipate noticeable continuum parameter differences after finding the best-fit absorber). Then we add a \textsc{xabs} component with a fixed ionisation parameter and turbulent velocity, at a specific (fixed) blueshift. These parameters are kept fixed to reduce the chance of running into a local minimum of the C-stat function during fitting and missing the global one. Afterwards, we fit for the column density of the absorber and recover the statistical fit improvement in \delcstat, similarly as in the Gaussian line scan case. The blueshift of the absorber is then changed according to a reasonably spaced grid (ideally again using the spectral resolution of the best instrument). The same search is repeated for a grid of ionisation parameters and a grid of turbulent velocities. This way we are sure we are not missing any parameter space inside our grids, albeit at a higher computational cost. A proxy to the significance of an absorber detection is again given by the \delcstat\ value, which does not take into account the number of trials. The real significance must be obtained by running Monte Carlo simulations where a fake spectrum is simulated using the continuum model as a template, and scanning it with the same physical model search. The resulting statistical significance is then 1 minus the ratio of occurences of fake lines (generated by Poisson noise) with the same or higher \delcstat\ fit improvement with respect to the total number of MC simulations. In this case, the Monte Carlo simulations are very computationally expensive, but they are accurate (see section \ref{Significance}).

The parameter space we search through is the following: the blueshift of the absorber between 0 and 120000 km/s, its ionisation parameter between 1.5 and 4.5, and its turbulent velocity between 250 and 5000 km/s. We choose the blueshift grid steps of 300 km/s, which oversamples slightly the RGS detector resolution, ionisation parameter steps of 0.1 and use 3 turbulence velocities: 250, 1000 and 5000 km/s. The results of the 1000 km/s turbulence velocity search are shown in Fig. \ref{Finaldata_XABS}. One can immediately notice the large fit improvement of about \delcstat$=$27.5 at $\sim$70000 km/s ($\sim$0.23c), for a scan with the ionisation parameter of $log~\xi=3.9$. The scans with other turbulent velocities do not show such large fit improvements at any systematic velocity and for any ionisation parameter.

\begin{figure*}
	\includegraphics[width=\textwidth]{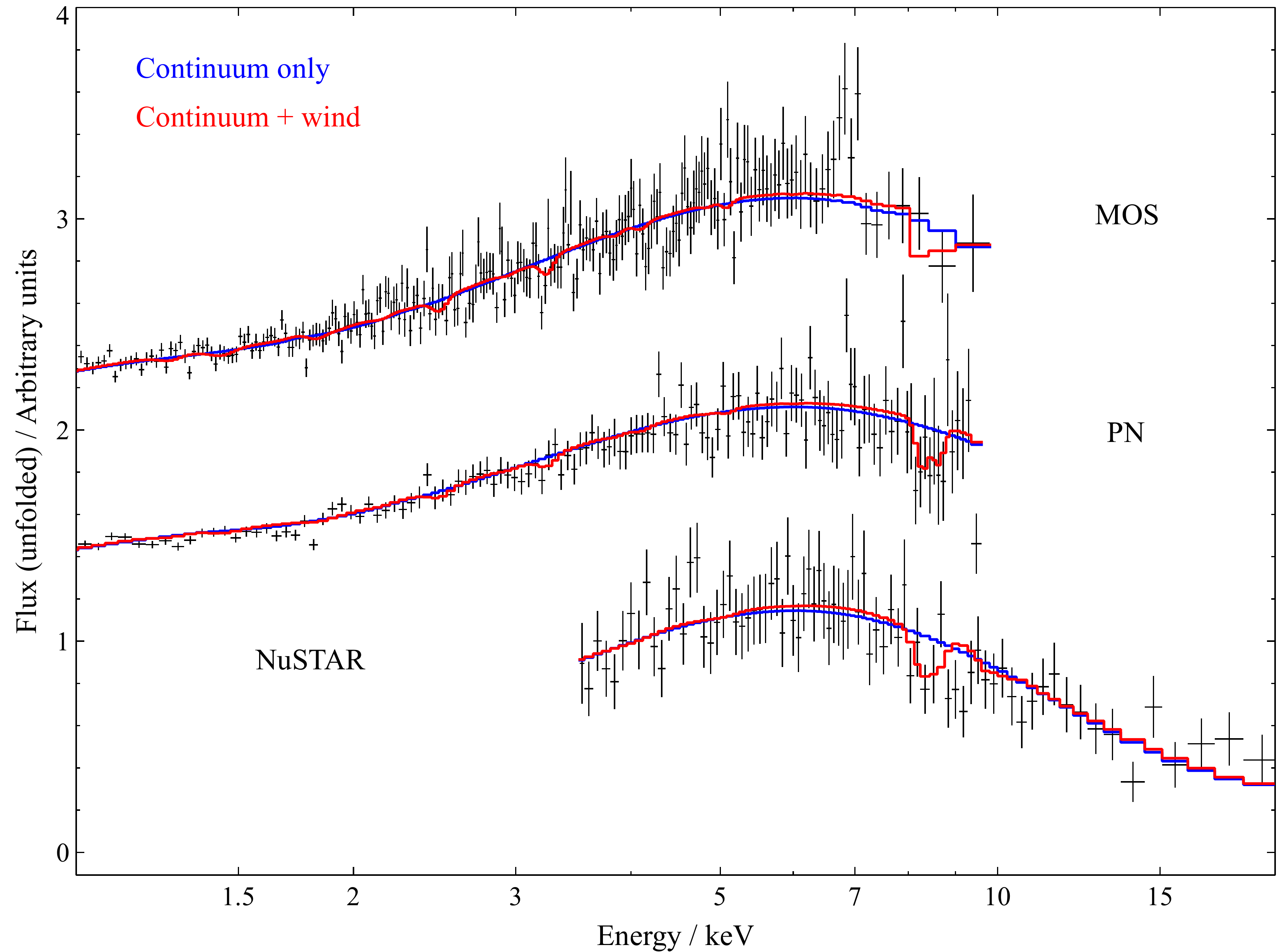}
    \caption{X-ray spectrum of the source during the second \xmm\ observation with simultaneous \nustar\ data. The Y-axis shows the source flux unfolded (with a constant), different instruments are shifted by constant amounts for plotting purposes. The X-axis shows energy in keV. The continuum only model is shown in blue colour, the continuum plus outflow model is in red. Data are stacked (MOS, FPM) and overbinned for plotting purposes.}
    \label{Finaldata_XABS_CCD}
\end{figure*}

\begin{figure}
	\includegraphics[width=\columnwidth]{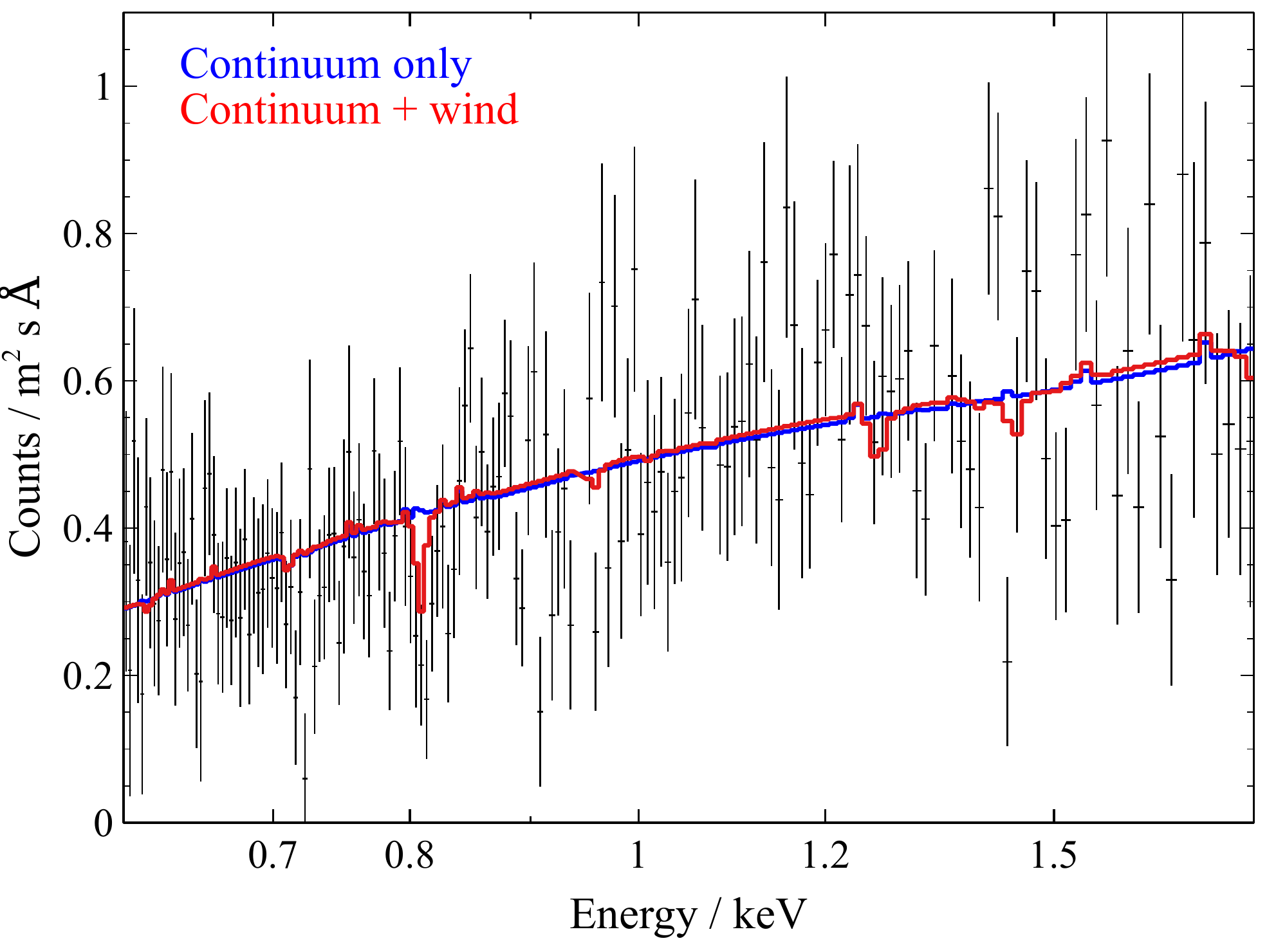}
    \caption{RGS spectrum during the second \xmm\ observation of the source. The Y-axis shows the source flux unfolded with a constant. The X-axis shows energy in keV. The continuum only model is shown in blue colour, the continuum plus outflow model is in red. Data from RGS1 and RGS2 detectors are stacked and overbinned for plotting purposes.}
    \label{Finaldata_XABS_RGS}
\end{figure}

We explore this solution in more detail by direct fitting in SPEX using all the available data. All the uncertainties are reported at 1$\sigma$ level. The best-fit solution shows the absorber outflowing at a projected (relativistically-corrected) speed of $65100^{+1000}_{-2200}$ km/s ($0.217^{+0.004}_{-0.007}~c$). Its column density is $1.2^{+1.9}_{-0.6}\times$10$^{23}$ \pcm, ionisation parameter is $log~\xi=3.92^{+0.19}_{-0.13}$ and turbulent velocity is $800^{+1100}_{-500}$ km/s. The continuum parameters are: powerlaw slope of $1.61^{+0.11}_{-0.13}$, blackbody temperatures $0.231^{+0.007}_{-0.008}$ keV and $2.73^{+0.08}_{-0.07}$ keV, and ISM column density of $(3.6 \pm 1.2)\times$10$^{20}$ \pcm. We find the statistical fit improvement of \delcstat$=$27.59. The spectrum fitted with the photoionized absorber is shown in Fig. \ref{Finaldata_XABS_CCD} (CCD data) and Fig. \ref{Finaldata_XABS_RGS} (RGS data). The strongest features of the absorber are iron absorption at 8-9 keV and oxygen absorption at 0.8 keV. We assume Solar abundances of elements \citep{Lodders+09} when fitting the \textsc{xabs} model.

To explore whether each detector agrees with this wind solution, we remove one instrument in turn from our analysis and perform the continuum and the continuum + wind spectral fit, calculating the \delcstat\ fit improvement between these 2 fits. Removing RGS data reduces our fit improvement upon adding the wind to broadband continuum to \delcstat=22.31. Removing PN data alone reduces the fit improvement to \delcstat=21.76, removing both MOS detectors decreases it to just \delcstat=17.44 and finally removing only FPM data decreases the fit improvement to \delcstat=21.33. The fact that removing any of the individual instruments from the analysis reduces the fit improvement \delcstat\ suggests that the wind signatures are present in all instruments, and hence they are not just Poisson noise.

The ultrafast wind is variable on relatively short timescales of tens of ks (see section \ref{firstxmmobs}) and it is possible that the wind is only present for a fraction of the observation (albeit the lightcurves do not show any significant flux change during the observation). To explore this possibility, we split the observation into halves by exposure time. Then we fit each half separately with the continuum and the continuum + wind model (as done with full data). We find that the wind signatures are slightly stronger in the second half of the observation, but the fit improvement is much lower due to decreased statistics. Hoping to increase the S/N ratio, we also cut out the first quarter and only fit the last 3 quarters of the observation exposure. This does not help either, the fit improvement is still weaker compared to the result achieved using the full dataset.

The MOS and PN spectra of the source also show hints of emission features, maybe with a P-Cygni profile. To see if they can be fitted with a consistent systematic velocity, we perform automated scans using physical emission models. We use a photoionised emitter model, \textsc{Photemis} in XSPEC, to represent emission from photoionised outflowing gas. Similarly, we use a collisionally ionised emitter model, \textsc{CIE} in SPEX, to describe emission from shocked gas. We run the automated scans in the same fashion as the photoionised absorber search. However, none of these show a significant statistical fit improvement. The maximum improvement obtained was \delcstat$\sim$14 only.

\subsection{Significance of the wind detection}
\label{Significance}

In the previous section, we show the possible detection of a relativistic outflow in the spectrum of NGC 300 ULX-1. However, it is not straightforward to rigorously assess its statistical significance. We obtained a relatively large \delcstat\ value of 27.59, which by itself would give a very high $\sigma$ significance, but it does not take into account the number of trials undertaken to find this solution. We will try to obtain the proper significance using 2 different approaches: with theory and with Monte Carlo simulations.

In theory, a search for spectral features on a noisy continuum can be viewed as follows: we have a number of free parameters in addition to broadband continuum, which we fit for in each step. In our case (section \ref{Phys_model_search}), there is just 1 variable: the column density of the absorber. The systematic velocity, the turbulent velocity and the ionisation parameter are all fixed in each fit, and then varied in a multi-parameter grid. Hence there is 1 additional degree of freedom compared to a continuum fit. Then we move the systematic velocity of the absorber and repeat the fit, and similarly we repeat the whole procedure in the grid of turbulent velocities and ionisation parameters. Effectively, we perform N trials with 1 degree of freedom for a presence of an absorber in our spectrum, where N is the number of points in our multi-parameter grid. This increases the null hypothesis probability of receiving a fit improvement, i.e. interpreting Poisson noise as a real detection.

To assess the null hypothesis probability, we perform the $\chi^{2}$ test. We group the RGS data by 20 counts per bin to achieve Gaussian-like statistics. For this part only, we stack the RGS detectors into a single spectrum to avoid overbinning any narrow spectral features. Data from the remaining instruments are already binned up so we can use $\chi^{2}$ statistics instead of C-stat when fitting. Repeating the same direct fit to all data as in the previous section, we achieve an even higher fit improvement of $\Delta\chi^{2}=29.68$ upon adding the photoionised absorber to broadband continuum. Our single trial p-value for such $\Delta\chi^{2}$ with 1 degree of freedom is about $5.1 \times 10^{-8}$ (obtained by integrating the $\chi^{2}$ probability density function). This is the probability of obtaining such fit improvement from spectral features introduced by noise only, if we performed just one measurement. We repeat the trial $N=401 \times 31 \times 3=37293$ times (the size of our multi-parameter grid - 120000 km/s velocity range divided into bins spaced by 300 km/s, ionisation parameters between 1.5 and 4.5 spaced by 0.1, and 3 different turbulent velocities), which brings up the null hypothesis probability to $(5.1 \times 10^{-8}) \times 37293=0.19$ per cent. The significance of our detection is therefore around 99.81 per cent or 3.1$\sigma$. We do stress, however, that this value needs to be taken with a grain of salt. It rather serves as a lower limit on the statistical significance as it assumes that all the trials were completely independent of each other, i.e. for example the finite instrument spectral resolution has no effect. Monte Carlo simulations will give a more realistic estimate.

We also perform Monte Carlo simulations to assess the significance of the wind detection in a rigorous way. We simulate a fake continuum spectrum using the real source continuum model (accounting for uncertainties in continuum parameters when simulating the spectrum). Ideally, afterwards we would perform the same automated search on fake data and count the number of wind detections stronger than the one found in real data. Unfortunately, it is not possible to do several thousands of simulations (to test a 3$\sigma$ significance) with proper error search, neither with the wind parameters (ionisation, turbulent velocity) being left free nor fixing them but having to create a grid of these parameters (as done with real data). Either of these approaches would require well above 100000 CPU hours. Making the systematic velocity grid significantly coarser is not an option either, since we could possibly miss a detection driven by the RGS data which has the highest spectral resolution. We instead devise a new approach, in which we separate the ionisation parameter space into 3 intervals, where $log~\xi$ is left free to vary: 1.5 to 2.5, 2.5 to 3.5 and 3.5 to 4.5, and free the turbulent velocity (but limit it to between 250 and 5000 km/s). Then we perform the automated search on the same simulated dataset in the 3 different $\xi$ intervals. Finally, the highest \delcstat\ fit improvement is recovered out of these 3 scans at each systematic velocity. This way we speed up the search process considerably, as well as reasonably assure no sensible parameter space is left unexplored (such as the fit falling into a local C-stat minimum and not discovering the global one). We made sure that this procedure finds the same features in real data as the original automated search. In the end, the required computational time to run 5000 Monte Carlo simulations with the wind model \textsc{xabs} is of the order of 10000 CPU hours (using a recent quad core i7 Intel CPU).

Out of the 5000 performed MC simulations, only one has a \delcstat\ fit improvement higher than 27.59, which was the highest \delcstat\ value found in the real data \textsc{xabs} search using both the grid and the sped-up 3 $\xi$ interval approach. This means, taking into account the total number of simulations, that the significance of the detection is around 99.98 per cent ($\sim$3.7$\sigma$), much higher than 3$\sigma$. For a 3$\sigma$ detection (corresponding to a false alarm probability of 0.27 per cent), the expected number of outliers equal or stronger than real data is 13. However, further simulations would be necessary to quantify the precise statistical significance. We note a good consistency between the result from the Monte Carlo simulations and our theoretical estimate which gives us a lower limit (3.1$\sigma$) on the significance simply given by the total number of trials performed by the automated search for photoionised absorber.

\subsection{The first XMM observation}
\label{firstxmmobs}

We extracted a spectrum of the object using all instruments during the first observation of \xmm\ (0791010101). A \nustar\ spectrum was also obtained for the same time interval. The flux of the source is very similar to its flux during the first observation, and the best-fit continuum spectral parameters are practically identical to those measured during the second observation (using the same continuum model). We perform the same Gaussian line scan and a physical wind search as we did with the second observation but find no signs of the outflow at all. There is no evidence for outflow neither at the same systematic velocity, nor at any similar velocity for any reasonable ionisation parameter $\xi$ and turbulent velocity $v$. The upper limit (1$\sigma$) on the column density of an absorber with the same ionisation, systematic and turbulent velocity as in the other observation is $0.2 \times 10^{23}$ \pcm.

Curiously, there seems to be a strong iron line at 6.7 keV in the PN data for this epoch, however it is not seen in either MOS nor FPM data. Note that this is the reverse of what is seen in the second observation, where an iron line is present in MOS data but not present in PN data.

\section{Discussion}
\label{sec:Discussion}

We performed an analysis of the simultaneous \xmm\ and \nustar\ data of the pulsating ultraluminous X-ray source NGC 300 ULX-1, which was taken in December 2016. We detect an ultrafast outflow in the spectrum of this object with a significance of more than 3$\sigma$. The fact that this significance is achieved by using 4 different instruments strengthens the credibility of the result. In addition, the statistical significance decreases upon removing any instrument from the analysis - which suggests that all the 4 different instruments individually prefer the wind solution. It is currently not possible to quantify the precise significance as this would require an extensive number of MC simulations. However, the current number of simulations performed already gives a very tight limit on the p-value of our result.

This is the first direct evidence of an ultrafast wind in a pulsating ULX. Also it is the first time that the evidence is seen in a ULX spectrum in both soft and hard X-ray detectors simultaneously, and fitted using a single physical model of the outflow. \citet{Walton+16} achieved a 3$\sigma$ detection of a possible iron K component to the wind in NGC 1313 X-1 \citep{Pinto+16}, but did not show a consistent fit of both soft and hard features at once.

The presence of a strong wind is consistent with the existence of super-Eddington accretion flow beyond the magnetosphere of the neutron star. The wind might be launched from this optically and geometrically thick part of the disk. If this is the case, the fact that the absorption is seen in both low- and high-energy X-ray data likely suggests that the absorption from this wind is imprinted on the emission from both the disk and the central accretion column \citep{Walton+18}.

Comparing with other ULXs with known outflows, the speed of 65000 km/s (about 0.22c) does not stand out as other ULXs have winds at 0.20-0.27c. A turbulent velocity of about 1000 km/s is also similar to what is observed in other objects. The ionisation parameter $\xi$, however, which is about $log~\xi=3.9 $~(erg~s$^{-1}$~cm), is higher than most best-fit solutions for wind in the 3 other ULXs. The only exception is NGC 1313 X-1, where an outflow with $log~\xi=4.5$ is allowed. In fact, the ionisation parameter of around $log~\xi=4$ is more similar to values measured in active galactic nuclei with known UFOs, e.g. IRAS 13224-3809 \citep{Parker+16, Pinto+18}.

The fact that we observe the wind in only one of the two observations suggests variability on relatively short timescales of tens of ks. \xmm\ observed the source on 2016-12-17, followed by a second observation on 2016-12-19. The time interval between the end of the first and the beginning of the second exposure was less than 10 hours. Such fast variability suggests a clumpy wind, in agreement with recent simulations \citep{Takeuchi+13, Kobayashi+18}. Alternatively, the viewing angle of the system might have changed to introduce obscuring material into the line of sight, however the change must have been small enough not to affect the broadband continuum, which is very similar during both observations. This seems unlikely. Wind variability has also been observed in NGC 1313 X-1 \citep[see Fig. 2 of][]{Pinto+16}.

Since we know the velocity and the ionisation parameter of the outflow, we can calculate its mechanical energy in comparison with the X-ray luminosity of the object, following the steps in \citet{Pinto+17}. The wind power can be expressed as $\dot{E}_{kin}=0.5\dot{M}u^{2}$ where $u$ is the wind velocity. The outflow rate $\dot{M}$ can then be determined using the definition of the ionisation parameter $\xi$: $\dot{M}=4\pi\Omega C_{V} L_{ion} m_{H} \mu u/\xi$ where $\Omega$ is the solid angle of the outflow as a fraction of $4\pi$, $L_{ion}$ is the ionising luminosity, and $C_{V}$ is the volume filling factor of the wind defining how clumpy it is. $m_{H}$ is the hydrogen (proton) mass and $\mu$ the mean atomic weight ($\sim$1.2 if taking solar abundances). We obtain the following for the mechanical power of the outflow: $\dot{E}_{kin}=2\pi\Omega C_{V} L_{ion} m_{H} \mu u^{3}/\xi$.

Substituting in the fitted and known variables, the kinetic power is $\dot{E}_{kin}\sim~$400$~\Omega C_{V} L_{ion}$. Assuming a fairly large solid angle fraction of $\Omega\sim0.5$ (wind at the inclination angles of between 40$^{\circ}$ and 80$^{\circ}$ from the rotation axis, as seen from the central object) and a volume filling factor of $C_{V}\sim0.3$ as found in simulations \citep{Takeuchi+13}, we reach $\dot{E}_{kin}\sim ~$60$~L_{ion}$. We can compare this value with the X-ray luminosity $L_{X}$ of the PULX. For an order of magnitude estimate, we assume that half of the X-ray luminosity ionises the wind, ignoring any possible strong beaming. Most of the ULX flux is usually found in the X-ray band so we do not need to account for any other ionising radiation. That means $L_{ion}/L_{X}\sim0.5$, which results in $\dot{E}_{kin}/L_{X}\sim~$30. Then taking $L_{X}\sim3 \times 10^{39}$~erg~s$^{-1}$ (obtained by fitting the X-ray spectrum), means that the total kinetic power is $\dot{E}_{kin}\sim 10^{41}$~erg~s$^{-1}$. This is an extreme value if compared to outflows in any sub-Eddington systems of stellar mass. However, it is similar (although on the higher end) to the values necessary to inflate ionised superbubbles many ULXs and microquasars are often located in, which require a mechanical inflating power of $10^{39}$ to several $10^{40}$~erg~s$^{-1}$ \citep{Pakull+02, Pakull+10, Cseh+12}. It is possible that in this case (a neutron star ULX), a larger fraction of the X-ray flux is beamed into the evacuated funnel and the ratio of the ionising luminosity $L_{ion}$ to the X-ray luminosity $L_{X}$ is much smaller, which would bring down the kinetic power $\dot{E}_{kin}$ to 10$^{39}$~-~10$^{40}$~erg~s$^{-1}$. Some level of anisotropy is in fact required for the system to pulse at all. In comparison, the kinetic energy needed to inflate the W50 bubble around the Galactic microquasar SS433 \citep[possibly a misaligned ULX, see][]{Fabrika+15} is a few times 10$^{39}$~erg~s$^{-1}$ \citep{Margon+84}. In this case, however, most of the kinetic energy is likely provided by its radio jets.

NGC 300 ULX-1 is one of two (the other one being NGC 7793 P13) currently known, easily observable neutron star ULXs. The two remaining PULXs are either too far or contaminated by close X-ray sources. The source is still active as of February 2018 at a luminosity of about $2 \times 10^{39}~$erg s$^{-1}$, and is a fascinating object, showing an ultrafast outflow appearing out of nothing on timescales of less than 100ks, in addition to pulsations with an extremely high pulsed fraction \citep{Carpano+18b}. We therefore strongly encourage further observations of the source with both high-resolution and broadband X-ray instruments.

The high ionisation factor of the ultrafast wind (log $\xi$=3.9) and its projected velocity of 0.24c means that most of its spectral features are in the >2 keV energy band. A majority of these features such as Mg, Si, S Ly$\alpha$ absorption lines is unresolvable with the current generation of X-ray instruments with the exception of Chandra gratings, which do not have sufficient collecting area (and therefore need long exposure times). The lines will however be easily resolved with the microcalorimeter onboard XARM \citep[Hitomi replacement,][]{Takahashi+10}, thus making it a prime instrument for studies of highly ionised outflows in ULXs.

\section{Conclusions}
\label{sec:Conclusions}

We performed a detailed, high spectral resolution and broadband search for an ultrafast outflow in the recently discovered pulsating ultraluminous X-ray source NGC 300 ULX-1. Our findings can be summarised as follows:

\begin{itemize}

\item{We found evidence for a UFO in this object with a significance of more than 3$\sigma$ during one of the two \xmm\ observations of the source, including simultaneous \nustar\ data. }

\item{Evidence for the UFO is seen in both soft X-ray data (driven by RGS gratings), and hard X-ray data (PN, MOS and FPM detectors). Removing any of the instruments from the simultaneous analysis reduces the total statistical significance.}

\item{The projected velocity of the outflow is $65100^{+1000}_{-2200}$ km/s ($0.217^{+0.004}_{-0.007}~c$), with a turbulent velocity of $800^{+1100}_{-500}$ km/s and a high ionisation factor of $3.92^{+0.19}_{-0.13}$. Its velocity is similar to outflows in other ULXs, but the ionisation parameter $\xi$ is more alike the outflows observed in active galactic nuclei.}

\item{Curiously, we find no signs of the wind during the other observation of the source by \xmm\ (with simultaneous \nustar\ coverage), which ended just 10 hours before the start of the second one. This could mean that either the wind is clumpy, or that the viewing angle of the system changed slightly, without affecting the overall broadband X-ray spectrum.}

\item{Further observations of the source are necessary to determine the process that drives the variability of the wind, and to place tighter constraints on its physical parameters as well as its interaction with the rest of the accretion flow in the system.}

\end{itemize}
\section*{Acknowledgements}

We are grateful to the anonymous referee for useful comments that improved the clarity and quality of the paper. PK acknowledges support from the STFC. CP and ACF acknowledge support from ERC Advanced Grant Feedback 340442. DJW acknowledges support from STFC Ernest Rutherford fellowships. This work is based on observations obtained with XMM-Newton, an ESA science mission funded by ESA Member States and USA (NASA). This research has made use of the NASA/IPAC Extragalactic Database (NED) which is operated by the Jet Propulsion Laboratory, California Institute of Technology, under contract with the National Aeronautics and Space Administration. This research has made use of the SIMBAD database, operated at CDS, Strasbourg, France.




\bibliographystyle{mnras}
\bibliography{References} 

\begin{thebibliography}{}
\makeatletter
\relax
\def\mn@urlcharsother{\let\do\@makeother \do\$\do\&\do\#\do\^\do\_\do\%\do\~}
\def\mn@doi{\begingroup\mn@urlcharsother \@ifnextchar [ {\mn@doi@}
  {\mn@doi@[]}}
\def\mn@doi@[#1]#2{\def\@tempa{#1}\ifx\@tempa\@empty \href
  {http://dx.doi.org/#2} {doi:#2}\else \href {http://dx.doi.org/#2} {#1}\fi
  \endgroup}
\def\mn@eprint#1#2{\mn@eprint@#1:#2::\@nil}
\def\mn@eprint@arXiv#1{\href {http://arxiv.org/abs/#1} {{\tt arXiv:#1}}}
\def\mn@eprint@dblp#1{\href {http://dblp.uni-trier.de/rec/bibtex/#1.xml}
  {dblp:#1}}
\def\mn@eprint@#1:#2:#3:#4\@nil{\def\@tempa {#1}\def\@tempb {#2}\def\@tempc
  {#3}\ifx \@tempc \@empty \let \@tempc \@tempb \let \@tempb \@tempa \fi \ifx
  \@tempb \@empty \def\@tempb {arXiv}\fi \@ifundefined
  {mn@eprint@\@tempb}{\@tempb:\@tempc}{\expandafter \expandafter \csname
  mn@eprint@\@tempb\endcsname \expandafter{\@tempc}}}

\bibitem[\protect\citeauthoryear{{Bachetti} et~al.,}{{Bachetti}
  et~al.}{2013}]{Bachetti+13}
{Bachetti} M.,  et~al., 2013, \mn@doi [\apj] {10.1088/0004-637X/778/2/163},
  \href {http://adsabs.harvard.edu/abs/2013ApJ...778..163B} {778, 163}

\bibitem[\protect\citeauthoryear{{Bachetti} et~al.,}{{Bachetti}
  et~al.}{2014}]{Bachetti+14}
{Bachetti} M.,  et~al., 2014, \mn@doi [\nat] {10.1038/nature13791}, \href
  {http://adsabs.harvard.edu/abs/2014Natur.514..202B} {514, 202}

\bibitem[\protect\citeauthoryear{{Binder}, {Williams}, {Kong}, {Gaetz},
  {Plucinsky}, {Dalcanton}  \& {Weisz}}{{Binder} et~al.}{2011}]{Binder+11}
{Binder} B.,  {Williams} B.~F.,  {Kong} A.~K.~H.,  {Gaetz} T.~J.,  {Plucinsky}
  P.~P.,  {Dalcanton} J.~J.,   {Weisz} D.~R.,  2011, \mn@doi [\apjl]
  {10.1088/2041-8205/739/2/L51}, \href
  {http://adsabs.harvard.edu/abs/2011ApJ...739L..51B} {739, L51}

\bibitem[\protect\citeauthoryear{{Brightman} et~al.,}{{Brightman}
  et~al.}{2018}]{Brightman+18}
{Brightman} M.,  et~al., 2018, \mn@doi [Nature Astronomy]
  {10.1038/s41550-018-0391-6}, \href
  {http://adsabs.harvard.edu/abs/2018NatAs...2..312B} {2, 312}

\bibitem[\protect\citeauthoryear{{Carpano}, {Haberl}, {Maitra}  \&
  {Vasilopoulos}}{{Carpano} et~al.}{2018a}]{Carpano+18b}
{Carpano} S.,  {Haberl} F.,  {Maitra} C.,   {Vasilopoulos} G.,  2018a, \mn@doi
  [\mnras] {10.1093/mnrasl/sly030}, \href
  {http://adsabs.harvard.edu/abs/2018MNRAS.476L..45C} {476, L45}

\bibitem[\protect\citeauthoryear{{Carpano}, {Haberl}  \& {Maitra}}{{Carpano}
  et~al.}{2018b}]{Carpano+18a}
{Carpano} S.,  {Haberl} F.,   {Maitra} C.,  2018b, The Astronomer's Telegram,
  \href {http://adsabs.harvard.edu/abs/2018ATel11158....1C} {11158}

\bibitem[\protect\citeauthoryear{{Cash}}{{Cash}}{1979}]{Cash+76}
{Cash} W.,  1979, \mn@doi [\apj] {10.1086/156922}, \href
  {http://adsabs.harvard.edu/abs/1979ApJ...228..939C} {228, 939}

\bibitem[\protect\citeauthoryear{{Cseh} et~al.,}{{Cseh} et~al.}{2012}]{Cseh+12}
{Cseh} D.,  et~al., 2012, \mn@doi [\apj] {10.1088/0004-637X/749/1/17}, \href
  {http://adsabs.harvard.edu/abs/2012ApJ...749...17C} {749, 17}

\bibitem[\protect\citeauthoryear{{Cseh} et~al.,}{{Cseh} et~al.}{2015}]{Cseh+15}
{Cseh} D.,  et~al., 2015, \mn@doi [\mnras] {10.1093/mnras/stv1308}, \href
  {http://adsabs.harvard.edu/abs/2015MNRAS.452...24C} {452, 24}

\bibitem[\protect\citeauthoryear{{Elias-Rosa}, {Mauerhan}  \& {van
  Dyk}}{{Elias-Rosa} et~al.}{2010}]{Elias-Rosa+10}
{Elias-Rosa} N.,  {Mauerhan} J.~C.,   {van Dyk} S.~D.,  2010, The Astronomer's
  Telegram, \href {http://adsabs.harvard.edu/abs/2010ATel.2636....1E} {2636}

\bibitem[\protect\citeauthoryear{{Fabrika}, {Ueda}, {Vinokurov}, {Sholukhova}
  \& {Shidatsu}}{{Fabrika} et~al.}{2015}]{Fabrika+15}
{Fabrika} S.,  {Ueda} Y.,  {Vinokurov} A.,  {Sholukhova} O.,   {Shidatsu} M.,
  2015, \mn@doi [Nature Physics] {10.1038/nphys3348}, \href
  {http://adsabs.harvard.edu/abs/2015NatPh..11..551F} {11, 551}

\bibitem[\protect\citeauthoryear{{Farrell}, {Webb}, {Barret}, {Godet}  \&
  {Rodrigues}}{{Farrell} et~al.}{2009}]{Farrell+09}
{Farrell} S.~A.,  {Webb} N.~A.,  {Barret} D.,  {Godet} O.,   {Rodrigues} J.~M.,
   2009, \mn@doi [\nat] {10.1038/nature08083}, \href
  {http://adsabs.harvard.edu/abs/2009Natur.460...73F} {460, 73}

\bibitem[\protect\citeauthoryear{{F{\"u}rst} et~al.,}{{F{\"u}rst}
  et~al.}{2016}]{Furst+16}
{F{\"u}rst} F.,  et~al., 2016, \mn@doi [\apjl] {10.3847/2041-8205/831/2/L14},
  \href {http://adsabs.harvard.edu/abs/2016ApJ...831L..14F} {831, L14}

\bibitem[\protect\citeauthoryear{{Gladstone}, {Roberts}  \& {Done}}{{Gladstone}
  et~al.}{2009}]{Gladstone+09}
{Gladstone} J.~C.,  {Roberts} T.~P.,   {Done} C.,  2009, \mn@doi [\mnras]
  {10.1111/j.1365-2966.2009.15123.x}, \href
  {http://adsabs.harvard.edu/abs/2009MNRAS.397.1836G} {397, 1836}

\bibitem[\protect\citeauthoryear{{Harrison} et~al.,}{{Harrison}
  et~al.}{2013}]{Harrison+13}
{Harrison} F.~A.,  et~al., 2013, \mn@doi [\apj] {10.1088/0004-637X/770/2/103},
  \href {http://adsabs.harvard.edu/abs/2013ApJ...770..103H} {770, 103}

\bibitem[\protect\citeauthoryear{{Israel} et~al.,}{{Israel}
  et~al.}{2017a}]{Israel+17a}
{Israel} G.~L.,  et~al., 2017a, \mn@doi [Science] {10.1126/science.aai8635},
  \href {http://adsabs.harvard.edu/abs/2017Sci...355..817I} {355, 817}

\bibitem[\protect\citeauthoryear{{Israel} et~al.,}{{Israel}
  et~al.}{2017b}]{Israel+17b}
{Israel} G.~L.,  et~al., 2017b, \mn@doi [\mnras] {10.1093/mnrasl/slw218}, \href
  {http://adsabs.harvard.edu/abs/2017MNRAS.466L..48I} {466, L48}

\bibitem[\protect\citeauthoryear{{Jansen} et~al.,}{{Jansen}
  et~al.}{2001}]{Jansen+01}
{Jansen} F.,  et~al., 2001, \mn@doi [\aap] {10.1051/0004-6361:20000036}, \href
  {http://adsabs.harvard.edu/abs/2001A%26A...365L...1J} {365, L1}

\bibitem[\protect\citeauthoryear{{Kaastra}, {Mewe}  \&
  {Nieuwenhuijzen}}{{Kaastra} et~al.}{1996}]{Kaastra+96}
{Kaastra} J.~S.,  {Mewe} R.,   {Nieuwenhuijzen} H.,  1996, in {Yamashita} K.,
  {Watanabe} T.,  eds, UV and X-ray Spectroscopy of Astrophysical and
  Laboratory Plasmas. pp 411--414

\bibitem[\protect\citeauthoryear{{Kaastra}, {Werner}, {Herder}, {Paerels}, {de
  Plaa}, {Rasmussen}  \& {de Vries}}{{Kaastra} et~al.}{2006}]{Kaastra+06}
{Kaastra} J.~S.,  {Werner} N.,  {Herder} J.~W.~A.~d.,  {Paerels} F.~B.~S.,  {de
  Plaa} J.,  {Rasmussen} A.~P.,   {de Vries} C.~P.,  2006, \mn@doi [\apj]
  {10.1086/507835}, \href {http://adsabs.harvard.edu/abs/2006ApJ...652..189K}
  {652, 189}

\bibitem[\protect\citeauthoryear{{Kalberla}, {Burton}, {Hartmann}, {Arnal},
  {Bajaja}, {Morras}  \& {P{\"o}ppel}}{{Kalberla} et~al.}{2005}]{Kalberla+05}
{Kalberla} P.~M.~W.,  {Burton} W.~B.,  {Hartmann} D.,  {Arnal} E.~M.,  {Bajaja}
  E.,  {Morras} R.,   {P{\"o}ppel} W.~G.~L.,  2005, \mn@doi [\aap]
  {10.1051/0004-6361:20041864}, \href
  {http://adsabs.harvard.edu/abs/2005A%26A...440..775K} {440, 775}

\bibitem[\protect\citeauthoryear{{Kennea}}{{Kennea}}{2018}]{Kennea+18}
{Kennea} J.~A.,  2018, The Astronomer's Telegram, \href
  {http://adsabs.harvard.edu/abs/2018ATel11229....1K} {11229}

\bibitem[\protect\citeauthoryear{{King}}{{King}}{2009}]{King+09}
{King} A.~R.,  2009, \mn@doi [\mnras] {10.1111/j.1745-3933.2008.00594.x}, \href
  {http://adsabs.harvard.edu/abs/2009MNRAS.393L..41K} {393, L41}

\bibitem[\protect\citeauthoryear{{Kobayashi}, {Ohsuga}, {Takahashi},
  {Kawashima}, {Asahina}, {Takeuchi}  \& {Mineshige}}{{Kobayashi}
  et~al.}{2018}]{Kobayashi+18}
{Kobayashi} H.,  {Ohsuga} K.,  {Takahashi} H.~R.,  {Kawashima} T.,  {Asahina}
  Y.,  {Takeuchi} S.,   {Mineshige} S.,  2018, \mn@doi [\pasj]
  {10.1093/pasj/psx157}, \href
  {http://adsabs.harvard.edu/abs/2018PASJ...70...22K} {70, 22}

\bibitem[\protect\citeauthoryear{{Kosec}, {Pinto}, {Fabian}  \&
  {Walton}}{{Kosec} et~al.}{2018}]{Kosec+18}
{Kosec} P.,  {Pinto} C.,  {Fabian} A.~C.,   {Walton} D.~J.,  2018, \mn@doi
  [\mnras] {10.1093/mnras/stx2695}, \href
  {http://adsabs.harvard.edu/abs/2018MNRAS.473.5680K} {473, 5680}

\bibitem[\protect\citeauthoryear{{Lau} et~al.,}{{Lau} et~al.}{2016}]{Lau+16}
{Lau} R.~M.,  et~al., 2016, \mn@doi [\apj] {10.3847/0004-637X/830/2/142}, \href
  {http://adsabs.harvard.edu/abs/2016ApJ...830..142L} {830, 142}

\bibitem[\protect\citeauthoryear{{Lodders}, {Palme}  \& {Gail}}{{Lodders}
  et~al.}{2009}]{Lodders+09}
{Lodders} K.,  {Palme} H.,   {Gail} H.-P.,  2009, \mn@doi [Landolt
  B{\"o}rnstein] {10.1007/978-3-540-88055-4_34}, \href
  {http://adsabs.harvard.edu/abs/2009LanB...4B...44L} {}

\bibitem[\protect\citeauthoryear{{Margon}}{{Margon}}{1984}]{Margon+84}
{Margon} B.,  1984, \mn@doi [\araa] {10.1146/annurev.aa.22.090184.002451},
  \href {http://adsabs.harvard.edu/abs/1984ARA%26A..22..507M} {22, 507}

\bibitem[\protect\citeauthoryear{{Middleton} et~al.,}{{Middleton}
  et~al.}{2013}]{Middleton+13}
{Middleton} M.~J.,  et~al., 2013, \mn@doi [\nat] {10.1038/nature11697}, \href
  {http://adsabs.harvard.edu/abs/2013Natur.493..187M} {493, 187}

\bibitem[\protect\citeauthoryear{{Middleton}, {Walton}, {Fabian}, {Roberts},
  {Heil}, {Pinto}, {Anderson}  \& {Sutton}}{{Middleton}
  et~al.}{2015}]{Middleton+15b}
{Middleton} M.~J.,  {Walton} D.~J.,  {Fabian} A.,  {Roberts} T.~P.,  {Heil} L.,
   {Pinto} C.,  {Anderson} G.,   {Sutton} A.,  2015, \mn@doi [\mnras]
  {10.1093/mnras/stv2214}, \href
  {http://adsabs.harvard.edu/abs/2015MNRAS.454.3134M} {454, 3134}

\bibitem[\protect\citeauthoryear{{Monard}}{{Monard}}{2010}]{Monard+10}
{Monard} L.~A.~G.,  2010, Central Bureau Electronic Telegrams, \href
  {http://adsabs.harvard.edu/abs/2010CBET.2289....1M} {2289}

\bibitem[\protect\citeauthoryear{{Mushtukov}, {Suleimanov}, {Tsygankov}  \&
  {Ingram}}{{Mushtukov} et~al.}{2017}]{Mushtukov+17}
{Mushtukov} A.~A.,  {Suleimanov} V.~F.,  {Tsygankov} S.~S.,   {Ingram} A.,
  2017, \mn@doi [\mnras] {10.1093/mnras/stx141}, \href
  {http://adsabs.harvard.edu/abs/2017MNRAS.467.1202M} {467, 1202}

\bibitem[\protect\citeauthoryear{{Narayan}, {Sadowski}  \& {Soria}}{{Narayan}
  et~al.}{2017}]{Narayan+17}
{Narayan} R.,  {Sadowski} A.,   {Soria} R.,  2017, \mn@doi [\mnras]
  {10.1093/mnras/stx1027}, \href
  {http://adsabs.harvard.edu/abs/2017MNRAS.469.2997N} {469, 2997}

\bibitem[\protect\citeauthoryear{{Pakull} \& {Mirioni}}{{Pakull} \&
  {Mirioni}}{2002}]{Pakull+02}
{Pakull} M.~W.,  {Mirioni} L.,  2002, ArXiv Astrophysics e-prints, \href
  {http://adsabs.harvard.edu/abs/2002astro.ph..2488P} {}

\bibitem[\protect\citeauthoryear{{Pakull}, {Soria}  \& {Motch}}{{Pakull}
  et~al.}{2010}]{Pakull+10}
{Pakull} M.~W.,  {Soria} R.,   {Motch} C.,  2010, \mn@doi [\nat]
  {10.1038/nature09168}, \href
  {http://adsabs.harvard.edu/abs/2010Natur.466..209P} {466, 209}

\bibitem[\protect\citeauthoryear{{Parker} et~al.,}{{Parker}
  et~al.}{2017a}]{Parker+17}
{Parker} M.~L.,  et~al., 2017a, \mn@doi [\mnras] {10.1093/mnras/stx945}, \href
  {http://adsabs.harvard.edu/abs/2017MNRAS.469.1553P} {469, 1553}

\bibitem[\protect\citeauthoryear{{Parker} et~al.,}{{Parker}
  et~al.}{2017b}]{Parker+16}
{Parker} M.~L.,  et~al., 2017b, \mn@doi [\nat] {10.1038/nature21385}, \href
  {http://adsabs.harvard.edu/abs/2017Natur.543...83P} {543, 83}

\bibitem[\protect\citeauthoryear{{Pinto}, {Middleton}  \& {Fabian}}{{Pinto}
  et~al.}{2016}]{Pinto+16}
{Pinto} C.,  {Middleton} M.~J.,   {Fabian} A.~C.,  2016, \mn@doi [\nat]
  {10.1038/nature17417}, \href
  {http://adsabs.harvard.edu/abs/2016Natur.533...64P} {533, 64}

\bibitem[\protect\citeauthoryear{{Pinto} et~al.,}{{Pinto}
  et~al.}{2017}]{Pinto+17}
{Pinto} C.,  et~al., 2017, \mn@doi [\mnras] {10.1093/mnras/stx641}, \href
  {http://adsabs.harvard.edu/abs/2017MNRAS.468.2865P} {468, 2865}

\bibitem[\protect\citeauthoryear{{Pinto} et~al.,}{{Pinto}
  et~al.}{2018}]{Pinto+18}
{Pinto} C.,  et~al., 2018, \mn@doi [\mnras] {10.1093/mnras/sty231}, \href
  {http://adsabs.harvard.edu/abs/2018MNRAS.tmp..231P} {}

\bibitem[\protect\citeauthoryear{{Pintore}, {Zampieri}, {Stella}, {Wolter},
  {Mereghetti}  \& {Israel}}{{Pintore} et~al.}{2017}]{Pintore+17}
{Pintore} F.,  {Zampieri} L.,  {Stella} L.,  {Wolter} A.,  {Mereghetti} S.,
  {Israel} G.~L.,  2017, \mn@doi [\apj] {10.3847/1538-4357/836/1/113}, \href
  {http://adsabs.harvard.edu/abs/2017ApJ...836..113P} {836, 113}

\bibitem[\protect\citeauthoryear{{Poutanen}, {Lipunova}, {Fabrika}, {Butkevich}
   \& {Abolmasov}}{{Poutanen} et~al.}{2007}]{Poutanen+07}
{Poutanen} J.,  {Lipunova} G.,  {Fabrika} S.,  {Butkevich} A.~G.,   {Abolmasov}
  P.,  2007, \mn@doi [\mnras] {10.1111/j.1365-2966.2007.11668.x}, \href
  {http://adsabs.harvard.edu/abs/2007MNRAS.377.1187P} {377, 1187}

\bibitem[\protect\citeauthoryear{{Shakura} \& {Sunyaev}}{{Shakura} \&
  {Sunyaev}}{1973}]{Shakura+73}
{Shakura} N.~I.,  {Sunyaev} R.~A.,  1973, \aap, \href
  {http://adsabs.harvard.edu/abs/1973A%26A....24..337S} {24, 337}

\bibitem[\protect\citeauthoryear{{Steenbrugge} et~al.,}{{Steenbrugge}
  et~al.}{2005}]{Steenbrugge+05}
{Steenbrugge} K.~C.,  et~al., 2005, \mn@doi [\aap]
  {10.1051/0004-6361:20047138}, \href
  {http://adsabs.harvard.edu/abs/2005A%26A...434..569S} {434, 569}

\bibitem[\protect\citeauthoryear{{Stobbart}, {Roberts}  \& {Wilms}}{{Stobbart}
  et~al.}{2006}]{Stobbart+06}
{Stobbart} A.-M.,  {Roberts} T.~P.,   {Wilms} J.,  2006, \mn@doi [\mnras]
  {10.1111/j.1365-2966.2006.10112.x}, \href
  {http://adsabs.harvard.edu/abs/2006MNRAS.368..397S} {368, 397}

\bibitem[\protect\citeauthoryear{{Str{\"u}der} et~al.,}{{Str{\"u}der}
  et~al.}{2001}]{Struder+01}
{Str{\"u}der} L.,  et~al., 2001, \mn@doi [\aap] {10.1051/0004-6361:20000066},
  \href {http://adsabs.harvard.edu/abs/2001A%26A...365L..18S} {365, L18}

\bibitem[\protect\citeauthoryear{{Takahashi} et~al.,}{{Takahashi}
  et~al.}{2010}]{Takahashi+10}
{Takahashi} T.,  et~al., 2010, in Space Telescopes and Instrumentation 2010:
  Ultraviolet to Gamma Ray. p. 77320Z (\mn@eprint {arXiv} {1010.4972}),
  \mn@doi{10.1117/12.857875}

\bibitem[\protect\citeauthoryear{{Takeuchi}, {Ohsuga}  \&
  {Mineshige}}{{Takeuchi} et~al.}{2013}]{Takeuchi+13}
{Takeuchi} S.,  {Ohsuga} K.,   {Mineshige} S.,  2013, \mn@doi [\pasj]
  {10.1093/pasj/65.4.88}, \href
  {http://adsabs.harvard.edu/abs/2013PASJ...65...88T} {65, 88}

\bibitem[\protect\citeauthoryear{{Turner} et~al.,}{{Turner}
  et~al.}{2001}]{Turner+01}
{Turner} M.~J.~L.,  et~al., 2001, \mn@doi [\aap] {10.1051/0004-6361:20000087},
  \href {http://adsabs.harvard.edu/abs/2001A%26A...365L..27T} {365, L27}

\bibitem[\protect\citeauthoryear{{Urquhart} \& {Soria}}{{Urquhart} \&
  {Soria}}{2016}]{Urquhart+16}
{Urquhart} R.,  {Soria} R.,  2016, \mn@doi [\mnras] {10.1093/mnras/stv2293},
  \href {http://adsabs.harvard.edu/abs/2016MNRAS.456.1859U} {456, 1859}

\bibitem[\protect\citeauthoryear{{Villar} et~al.,}{{Villar}
  et~al.}{2016}]{Villar+16}
{Villar} V.~A.,  et~al., 2016, \mn@doi [\apj] {10.3847/0004-637X/830/1/11},
  \href {http://adsabs.harvard.edu/abs/2016ApJ...830...11V} {830, 11}

\bibitem[\protect\citeauthoryear{{Walton} et~al.,}{{Walton}
  et~al.}{2014}]{Walton+14}
{Walton} D.~J.,  et~al., 2014, \mn@doi [\apj] {10.1088/0004-637X/793/1/21},
  \href {http://adsabs.harvard.edu/abs/2014ApJ...793...21W} {793, 21}

\bibitem[\protect\citeauthoryear{{Walton} et~al.,}{{Walton}
  et~al.}{2016a}]{Walton+16cyg}
{Walton} D.~J.,  et~al., 2016a, \mn@doi [\apj] {10.3847/0004-637X/826/1/87},
  \href {http://adsabs.harvard.edu/abs/2016ApJ...826...87W} {826, 87}

\bibitem[\protect\citeauthoryear{{Walton} et~al.,}{{Walton}
  et~al.}{2016b}]{Walton+16}
{Walton} D.~J.,  et~al., 2016b, \mn@doi [\apjl] {10.3847/2041-8205/826/2/L26},
  \href {http://adsabs.harvard.edu/abs/2016ApJ...826L..26W} {826, L26}

\bibitem[\protect\citeauthoryear{{Walton} et~al.,}{{Walton}
  et~al.}{2018a}]{Walton+18}
{Walton} D.~J.,  et~al., 2018a, \mn@doi [\mnras] {10.1093/mnras/stx2650}, \href
  {http://adsabs.harvard.edu/abs/2018MNRAS.473.4360W} {473, 4360}

\bibitem[\protect\citeauthoryear{{Walton} et~al.,}{{Walton}
  et~al.}{2018b}]{Walton+18b}
{Walton} D.~J.,  et~al., 2018b, \mn@doi [\apjl] {10.3847/2041-8213/aabadc},
  \href {http://adsabs.harvard.edu/abs/2018ApJ...857L...3W} {857, L3}

\bibitem[\protect\citeauthoryear{{Webb} et~al.,}{{Webb} et~al.}{2012}]{Webb+12}
{Webb} N.,  et~al., 2012, \mn@doi [Science] {10.1126/science.1222779}, \href
  {http://adsabs.harvard.edu/abs/2012Sci...337..554W} {337, 554}

\bibitem[\protect\citeauthoryear{{den Herder} et~al.,}{{den Herder}
  et~al.}{2001}]{denHerder+01}
{den Herder} J.~W.,  et~al., 2001, \mn@doi [\aap] {10.1051/0004-6361:20000058},
  \href {http://adsabs.harvard.edu/abs/2001A%26A...365L...7D} {365, L7}

\makeatother
\end{thebibliography}




%
%


\bsp	
\label{lastpage}
\end{document}